**Chiral Interactions of Histidine in a Hydrated Vermiculite Clay**

Donald G. Fraser,[1*] H. Christopher Greenwell,[2] Neal T. Skipper,[3] Martin V. Smalley,[4] Michael A. Wilkinson,[3,5] Bruno Demé,[5] R. K. Heenan.[6]

[1]*Department of Earth Sciences, University of Oxford, Parks Road, Oxford OX1 3PR, UK*

[2]*Department of Chemistry, Durham University, South Road, Durham DH1 3LE, UK*

[3]*Department of Physics & Astronomy, University College London, Gower Street, London WC1E 6BT, UK*

[4]*Department of Physics, University of York, Heslington, York YO10 5DD, UK*

[5]*Institut Laue Langevin, F-38042 Grenoble, Cedex 9, France*

[6]*ISIS Facility, R3 1-22, Rutherford Appleton Lab, HSIC, Didcot, OX11 0QX, UK*

*Email: don@earth.ox.ac.uk*

Graphical Contents Entry

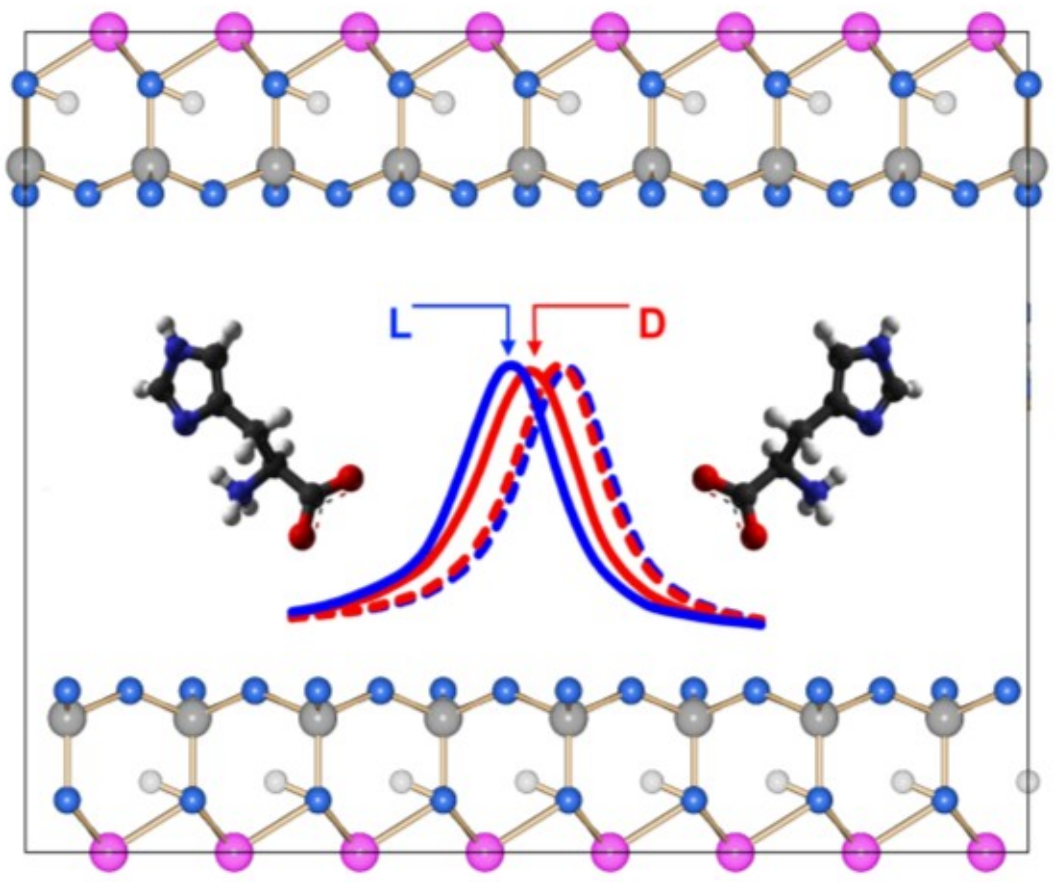

We report chiral differences in the effects of D- and L-histidine on the interlayer spacing of vermiculite clays that are relevant in assessing the role of clays in the origin of bio-homochirality.

Abstract

Recent work suggests a link between chiral asymmetry in the amino acid iso-valine extracted from the Murchison meteorite and the extent of hydrous alteration.
We present the results of neutron scattering experiments on an exchanged, 1-dimensionally ordered n-propyl ammonium vermiculite clay. The vermiculite gel has a (001) d-spacing of order 5nm at the temperature and concentration of the experiments and the d-spacing responds sensitively to changes in concentration, temperature and electronic environment. The data show that isothermal addition of D-histidine or L-histidine solutions produces shifts in the d-spacing that are different for each enantiomer. This chiral specificity is of interest for the question of whether clays could have played an important role in the origin of biohomochirality.

## Introduction

Understanding the interaction of biomolecules such as amino acids and simple sugars with hydrated mineral surfaces is key to unraveling the origins of life on Earth[1]. The chirality of some crystalline mineral phases, such as quartz and calcite, has long been proposed as a possible mechanism for selectively determining the production of particular prebiotic enantiomers[2]. However, neither calcite nor quartz has been shown to have any heterogeneous catalytic function in aiding biomolecule polymerization. Fundamental questions arise as to how the first organic molecules arrived on Earth and were built up into complex molecules through prebiotic chemistry, whilst being protected from the harsh environment present at the time. Though the view that life arose through an "RNA World" is widely supported[3-4], there is also an alternative hypothesis that proteins originated first on the early Earth[5]. Lambert has recently reviewed the organization and polymerization of amino acids at mineral surfaces[6].

Clay minerals have been shown to be efficient at concentrating amino acids from solution[7-9], especially those with charged side groups, such as lysine or histidine. Clays have also been shown to catalyse peptide bond formation in their own right[10-12], and when used in conjunction with salt induced peptide formation (SIPF) chemistries[13-15].

Clay minerals in carbonaceous meteorites have also been shown to be of potential importance for protecting organic molecules from oxidizing conditions[16]. At least 70 chiral amino acids have been identified in meteorites. These were believed to show almost no chiral selectivity, being close to 50:50 racemic mixtures[17], but recently an important chiral asymmetry in iso-valine extracted from the Murchison meteorite has been reported[18]. Laboratory gas condensation experiments also generate amino acids, though, again, these show almost no chiral selectivity[19]. In contrast, amino acids in proteins and enzymes in organisms are usually 100% L-enantio-specific.

Prebiotic chemistry must address the question of homochirality and the emergence of only L-amino acids in virtually all living organisms. Goldberg showed, once seeded with one enantiomer, clay surfaces enhance selectivity for that enantiomer during crystallization cycles[20]. Recent experiments have pointed towards chirally selective adsorption and reactivity in clay mineral suspensions[21-22]. Anionic clays have also been shown to have kinetic chiral selectivity for histidine[23]. The anionic clays have much higher charge density and packing than for the cationic systems. Yu *et al.* used density functional theory computer simulations to examine the energy of L- and D-alanine on the surface of nontronite clay, finding a preference by some 25 kJ/mol for the L- form[23]. Adsorption of amino acids at clay mineral surfaces is also of interest in a range of natural phenomena[24-26]. Industrially, amino acid chiral selectivity and racemization stabilization effects have been observed in anionic clays (layered double hydroxides)[27].

In many adsorption studies, structural analysis of the clay-amino acid system is often conducted post reaction, whereas there are no high-resolution *in situ* structural studies of fully hydrated systems. In the work presented here, we studied the addition of D- and L-histidine to *n*-propylammonium vermiculite gels (with layer spacing ~50 Å). The alkylammonium vermiculite gels have been established as a model system for studying clay swelling and colloid stability[28]. The swelling is nearly perfectly one-dimensional, taking place perpendicular to the plane of the silicate layers, and leads to the formation of gels with *d*-values between 40 and 90 Å along the swelling axis. Even for the largest expansions, this *d*-value remains sufficiently well defined to be measured by small-angle neutron diffraction (SANS).

## Experimental

Crystals of natural Eucatex vermiculite of dry composition $Si_{6.13}Mg_{5.44}Al_{1.65}Fe_{0.50}Ti_{0.13}Ca_{0.13}Cr_{0.01}K_{0.01}O_{20}(OH)_4Na_{1.29}$ were exchanged in $CH_3CH_2CH_2NH_2{\cdot}HCl$ (Sigma-Aldrich 242543, >99%) solution over a period of several years. Around two weeks before the experiments crystals of approximate dimensions 10 × 10 × 2 mm were placed in stock solution of propylammonium chloride in $D_2O$ at the required concentration, and were equilibrated for the neutron scattering experiments with regular exchange of supernatant solution. Matched solutions containing D-histidine (Puriss® 53321, ≥99%) or L-histidine (ReagentPlus® H8000, ≥99% both analysed and certificated by TLC) were made by addition of reagent to $D_2O$ or identical stock solution of propylammonium chloride.

Individual equilibrated gel crystals were selected from solution and placed in sample containers. The samples were then bathed in solution as detailed below. The swollen gels are extremely soft and delicate, and all handling was conducted under solution.

Previous studies of these gels have established the relationship between *d*-spacing and concentration in propylammonium chloride solution[28]. In the region of most interest, around 0.35M, the equilibrium *d*-spacing increases by around 1 Å for a 0.01M decrease in concentration. The changes we report here are therefore significant when compared with the reliability of the solution concentrations.

Preliminary experiments were conducted on the LOQ instrument at the ISIS Pulsed Neutron source, Rutherford Appleton Laboratory. We equilibrated stocks of gels in 0.1M, 0.2M and 0.35M propylammonium chloride in $D_2O$. At these concentrations the *d*-spacings are approximately 110, 75 and 50 Å respectively. As far as possible gels were selected from the stock as matched pairs, and when both showed a clear

(001) Bragg peak we then exposed one to L- and one to D-histidine. In this way we studied 6 pairs in total; 1 at 0.1M, 2 at 0.2M and 3 at 0.36M.

Samples were placed flat on the bottom of sealed 10 × 10 mm Hellma QS cells and were bathed in $3cm^3$ of the supernatant solution of the required concentration. The sample height and alignment was adjusted individually by laser beam, and the small angle neutron scattering was measured in the $Q$ range 0.006 - 0.24 $Å^{-1}$. After this initial characterisation the samples were exposed to amino acid by the addition of $0.5cm^3$ of 0.18M L- or D-histidine in $D_2O$. After approximately 1 hour the neutron scattering was measured again. Temperature was maintained at 30 ±1°C throughout by a water bath in thermal contact with the sample block. Cycles through a temperature range were conducted to confirm the crystal-gel transition.

Scattering intensity was corrected for background from the empty container and normalised to an absolute scale by reference to a water standard and a beam transmission measurement. Due to the sample orientation the Bragg scattering from the (00*l*) reflections was recorded in the top and bottom quadrants of the area detector. The data were binned in $Q$ with a constant bin-width of 0.001 $Å^{-1}$. The scattered intensities were fitted with the program *Fish* [Heenan, R. K. *Fish data analysis program*, RAL-Report-89-129, Rutherford Appleton Laboratory (UK) **1989**] using Lorentzian Bragg profiles with a form factor derived for a distribution of oriented sheets[29-30]. Example fits are shown in Figure 1.

In all cases the *d*-spacing was <u>increased</u> at constant concentration by exposure to histidine. If we express this expansion in terms of the fractional change $d/d_0$ then the average shift is 1.117 ±0.020 for L-histidine and 1.102 ±0.020 for D-histidine. The average expansion was therefore 1.5% greater for L-histidine than D-histidine, but tantalisingly, this is on the limit of the statistical precision of the experiments.

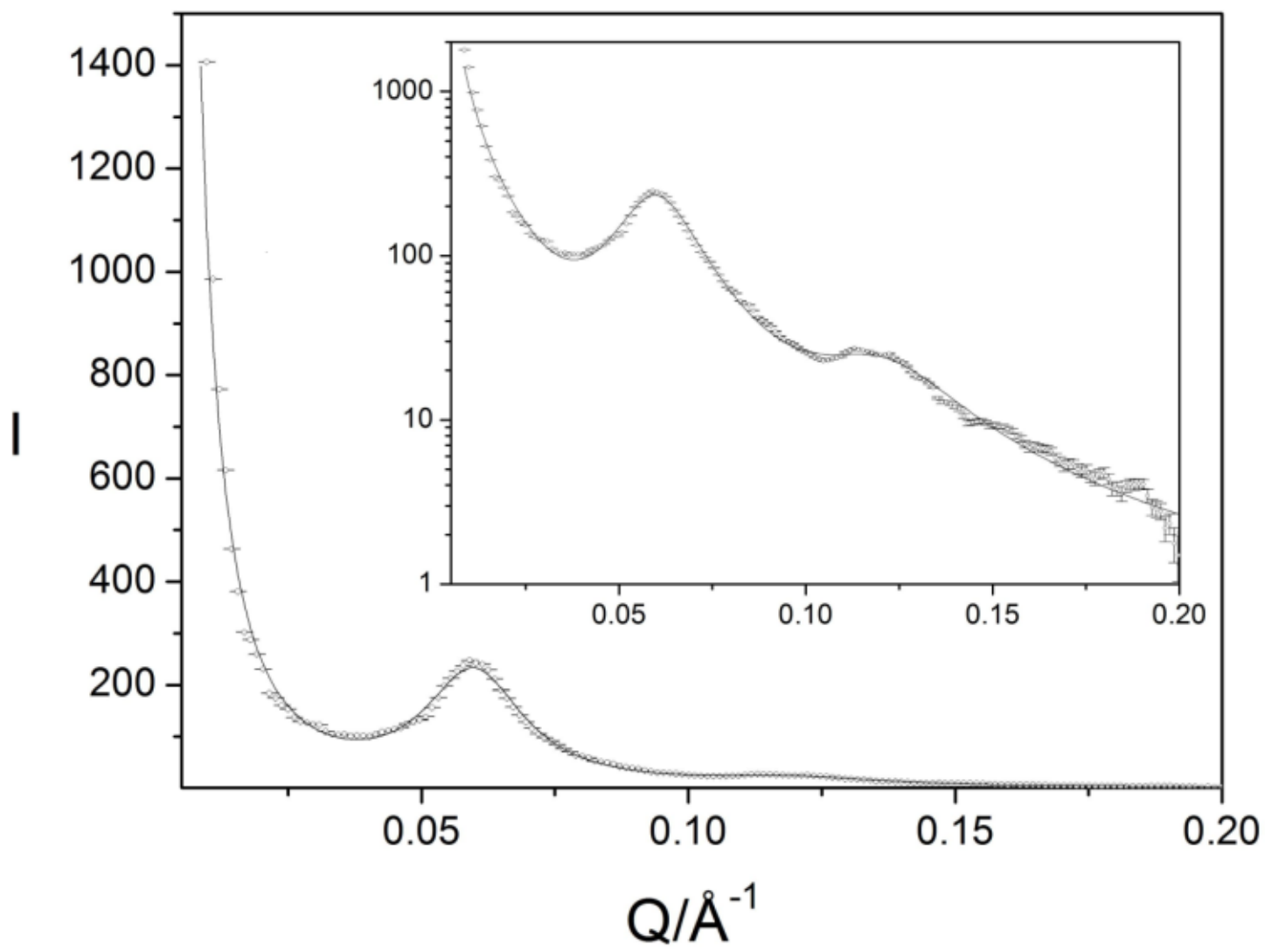

Figure 1: Example of neutron scattering intensity I (arbitrary units) obtained from the instrument LOQ for the sample exposed to 0.1M propyl ammonium chloride before exposure to L-histidine. Data (+) and fit (line). Inset shows intensity I on a $\log_{10}$ scale.

Following these initial studies on the LOQ instrument at ISIS we exploited the increased Q-range of station D16 at the Institut Laue-Langevin, Grenoble, to measure 1$^{st}$ and higher order (00*l*) peaks in both the crystalline and gel phases of propylammonium ion substituted vermiculite clay. The instrument was operated at a constant wavelength of 4.728 Å with a pixelated detector giving a *Q*-range of 0.04 – 0.5 Å$^{-1}$ with a bin-width of 0.0025 Å$^{-1}$. The detector efficiency was normalized by reference to a water standard.  Samples of gel phase propylammonium vermiculite were prepared via the method of Williams *et al*.[31-32], by soaking crystals in a deuterated solution of 0.34M propylammonium chloride. We first measured the structure of the gels in the absence of amino acid as a control. Clay gels were equilibrated in 0.3387M propylammonium chloride in $D_2O$ and a high-quality sample was cleaved into two halves (Figure 2). Each half was then placed in a QS cell with 4 × 10 mm sample space under supernatant solution and was aligned using a laser beam and then via a rocking curve centred on the (001) Bragg reflection. Temperature was maintained at 25.5 ±0.5°C via a thermal bath connected to a neutron-transparent thermal jacket.

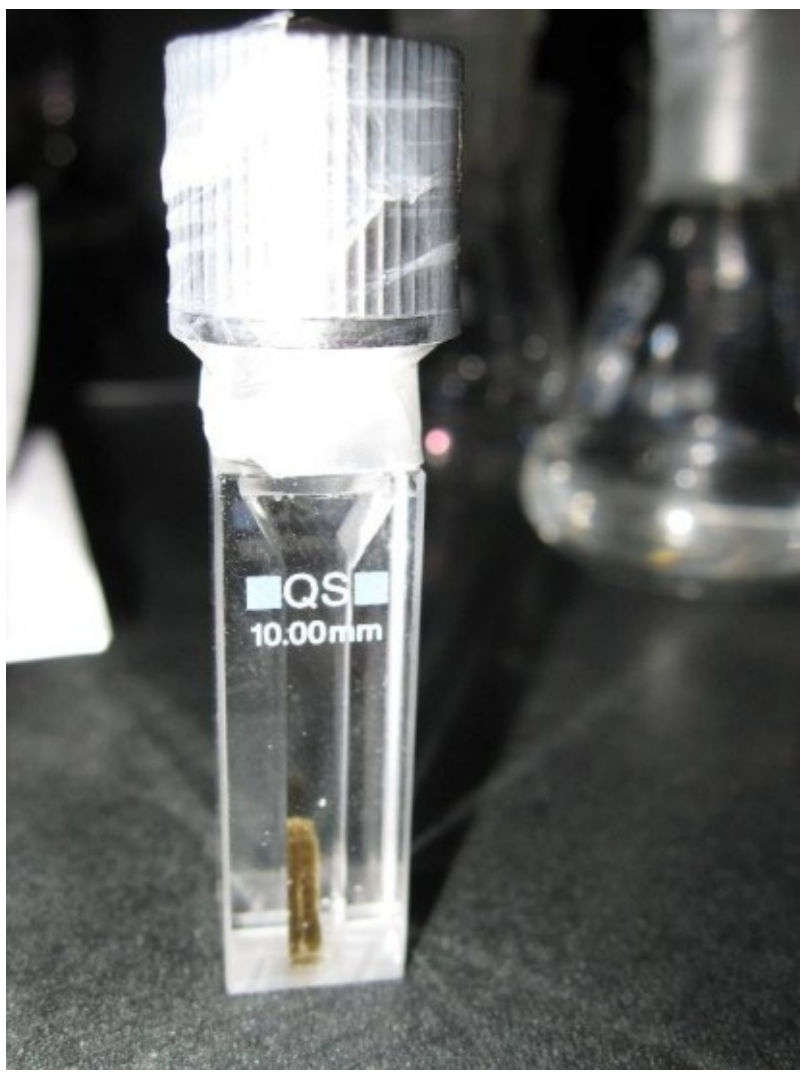

Figure 2. Vermiculite clay gel in solution for measurement on D16.

After measurement on the samples under pure 0.3387M propylammonium chloride in $D_2O$, the solution was replaced with 0.3387m propylammonium chloride containing either 0.05M L- or D-histidine in $D_2O$. After 5 full exchanges of solution over the samples, conducted over a period of around 6 hours during which the neutron scattering profile was monitored, steady state equilibrium was achieved. The final profile was then measured.

The normalized intensities were fitted as before using the program *Fish* using Lorentzian Bragg profiles with a form factor derived for a distribution of oriented sheets. A typical fit is shown in Figure 3.

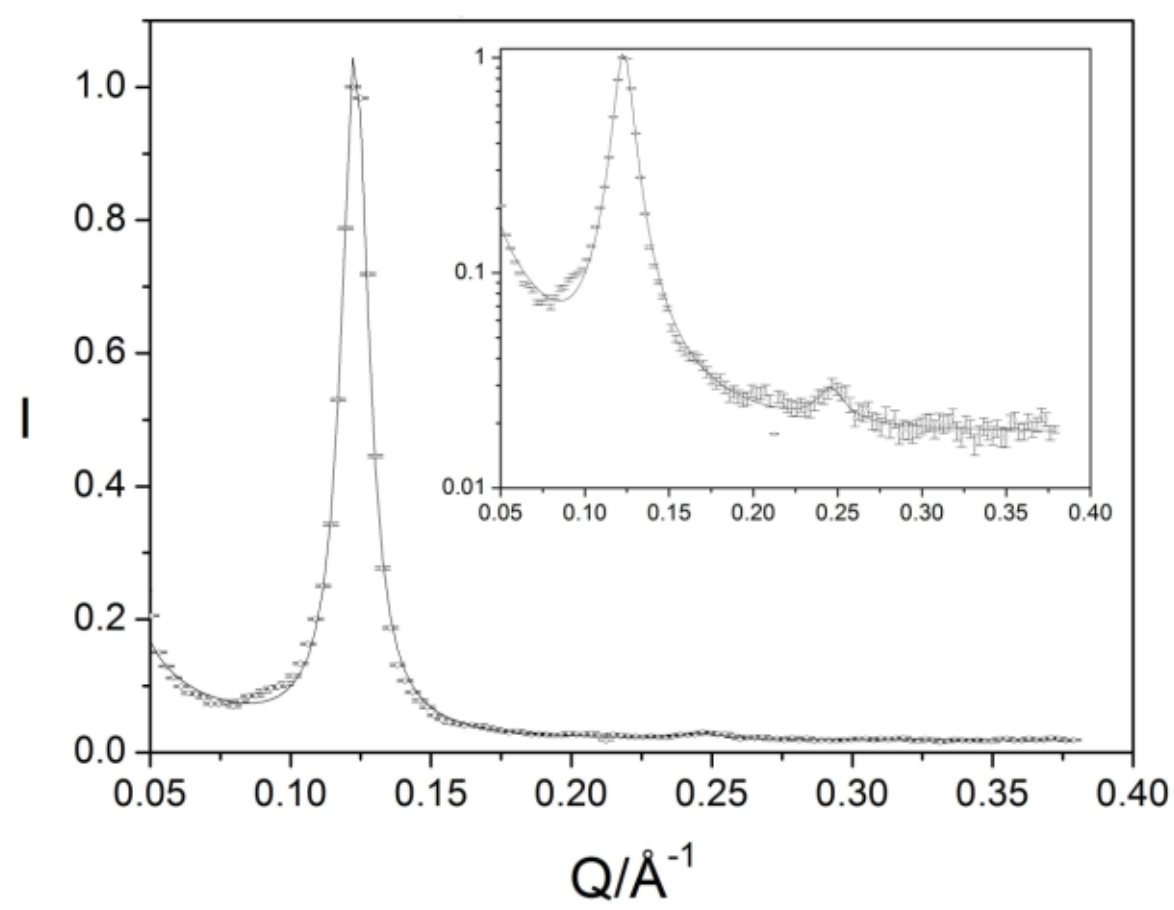

Figure 3: Example of neutron scattering intensity I (arbitrary units) obtained from the instrument D16 for the sample exposed to 0.3387M propyl ammonium chloride and 0.05M L-histidine. Data (+) and fit (line). Inset shows intensity I on a $\log_{10}$ scale.

## Discussion

It is known that the phase transition between these two phases occurs at around 310K[28]. Our hypothesis was that the thermodynamics of the phase diagram and the interlayer *d*-spacings of the clay, would be sensitive to the adsorption of histidine enantiomers onto the clay surface, and that we should therefore be able detect their uptake/location and any preferential chiral interaction. Figure 4 shows data obtained for the two nearly identical halves of untreated gel measured first in pure 0.34M propylammonium solution. These can be compared in the same Figure with the data for the same vermiculite samples measured after 5 exchanges in L- and D-histidine respectively.

The peak in all cases arises from the (001) Bragg reflection – and therefore directly gives the *d*-spacing of the sample. Increase in ordinary solute concentration would cayuse the d-spacing to contract. The data in Figure 4 show clearly that the amino acids act "anti-osmotically" so that addition of amino acid, and hence increase in the overall solute concentration, causes the *d*-spacing to widen relative to that of the pure control clay system. This suggests that histidine adsorbs to the clay surface, screening the Coulombic charge. In addition, there is evidence of a chiral shift: L-histidine increases the *d*-spacing relative to D-histidine. The value for the control was 48.78Å; D-histidine shifted from 49.06 to 50.33Å; L-histidine shifted from 48.76 to 51.11Å. The ~1.5% difference between L- and D- is significant in terms of the accuracy of the instrument, and is also consistent with our previous experiments on LOQ.

To provide further information on the nature of the amino acid adsorption, a separate experiment was conducted in which a previously L-histidine-equilibrated sample was exposed to D-histidine. In this case there was no significant reverse shift in the *d*-spacing on going from L- to D- over the timescale of the experiment (~ 6 hours). This lack of short term reversibility lends support to the notion that the amino acids are bound directly and strongly to the clay surface.

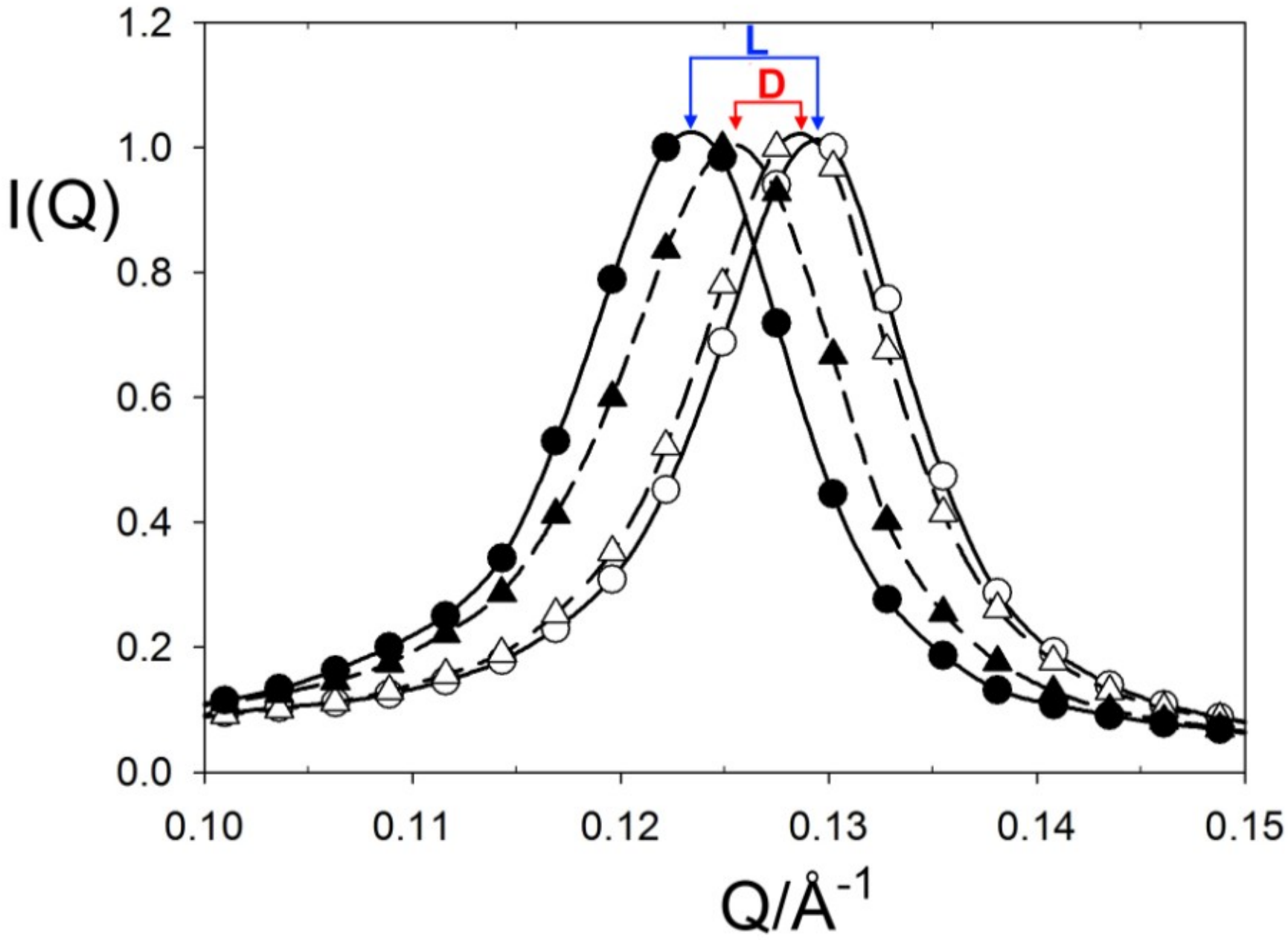

***Figure 4.*** Shift in *d*-spacing of histidine intercalated propyl-ammonium vermiculite clay gels and control systems. L-histidine and control: filled and empty circles. D-histidine and control: filled and empty triangles.

The behaviour of the histidine molecules in the inter-layer region of the hydrated amino acid clay system was modeled by Monte-Carlo and molecular dynamics simulations and a schematic of the interlayer region of the hydrated clay is shown in Fig.5. The Monte Carlo algorithm Monte was used to generate the initial coordinates for a hydrated propylammonium vermiculite with phenol in the interlayer. Forcefield parameters were the OPLS. The equilibrated system was taken and the phenol molecules converted to histidine. 40 ps of molecular dynamic simulation was then run using the Dreiding forcefield, with the charges previously assigned, with the Forcite code in the Accelrys Inc Materials Studio. It is apparent that a large proportion of the histidine molecules are indeed adsorbed at the mineral surfaces as suggested above.

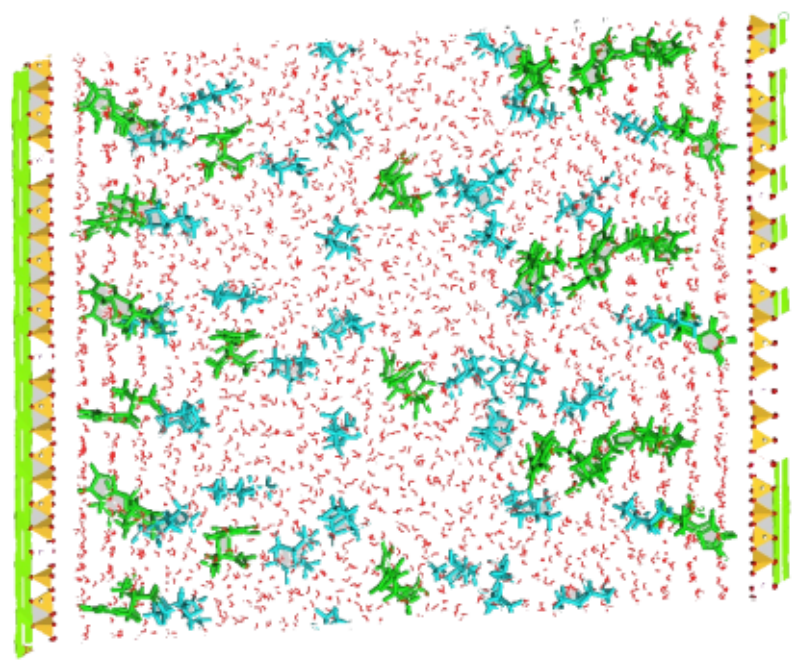

***Figure 5.*** Schematic based on molecular simulations showing interlayer arrangement of histidine (green coloured) molecules in a propylammonium (blue molecules) vermiculite clay (color: Si = orange; Al = magenta; Mg = green; O = red; H = white).

To try to determine whether the effects described above were due to preferential uptake of one isomer over the other, chiral high performance liquid chromatography (HPLC) was undertaken. Experiments were carried out in 8 ml screwtop glass vials. A 250ml stock solution of 0.40mg/ml DL histidine and 0.338M $PrNH_3Cl$ was prepared. Raw Eucatex vermiculite, which had been stored under distilled water, was dried at 80 ˚C for 24 hours. Sodium vermiculite was exchanged twice over 24 hours with 0.388M propylamine solution, and dried at 140 ˚C for 1 hour. Different weights of clay were then added to a constant amount of histidine solution, so that the change in histidine concentration could be found. The raw vermiculite samples were left to equilibrate over 24 hours, whilst the propylamine vermiculite equilibrated over 1.5 hours.

*Analysis method* A Perkin-Elmer Peltier plate HPLC with autosampler, in the Analytical Laboratory, Department of Chemistry, Durham University, was used for all analyses. A Crownpak CR(+) chiral HPLC column (D form eluted first) was used throughout. Stock 20% perchloric acid diluted to give pH 1 was used as an eluent, with the histidine underivatised. A temperature of 10 °C and a flow rate of 0.1 $ml.min^{-1}$ was found to give good separation and could be reliably maintained. An UV detector, λ 200 nm, was used. The retention times were 4.73 min and 6.27 min for the D- and L-histidine, respectively.
A set of standards was made up at the experimental concentration of 0.388M and with the change in enantiomeric excess (*ee*) set at 5.00%, 1.00%, 0.50% and 0.05%. For a mixture of isomers the relative area under each peak, expressed as a percentage of the total of the two isomer peaks, indicates the enantiomeric excess. At 5% and 1% *ee* the steps were reasonably well resolved. The relative high values of the area measured (in $\mu V.\ s^{-1}$) lead to a spread in the results at the lower *ee* of 0.5% and 0.05%. However, when these are converted to percentage *ee*, a clearer trend was defined. Even at 0.5% *ee*, the observed changes seem valid. In summary, it would seem that the instrument and method used were capable of discerning, within reasonable error, increments down to 0.05%, though the absolute value may be slightly different to the standard.

To investigate the effects of adsorption on the vermiculite clay varying weights of clay were added to the racemic mix of D/L histidine and the supernatant measured for preferential adsorption of amino acid. In addition to testing propylammonium exchanged clay as used in the neutron scattering experiments, the sodium clay and propylammonium solution were tested, and the sodium clay alone was tested to determine any effect the propyl ammonium clay / solution may have had.

Within the error present within the data, and allowing for the slight non-equivalence in the baseline solutions with no clay present, it is not possible to discern an *ee* in favour of either enantiomer. The size of the variation in the data was +/- 0.3 % *ee* .

Thus, within the experimental error, we could find no evidence for preferential intercalation of either enantiomer. This suggests that the observed shift in d-spacing is purely due to subtle electrostatic screening effects between the two isomers.

In conclusion, we demonstrated that the effect of different enantiomers of an amino acid may alter physical and structural properties of clay systems, at low amino acid concentrations and over a long length-range. This considerably extends the previous studies of chiral selectivity by clays, which usually require high packing densities and low hydration states for differences to become apparent. HPLC analysis was unable to discern any compositional differences between the different systems, indicating that the observed effect is due to structural differences within the interlayer.

Acknowledgements. We are grateful to ISIS and ILL for neutron beamtime. We thank Dr A. Congreve, Department of Chemistry, Durham University, for assistance with HPLC measurements. DGF gratefully acknowledges financial support from Worcester College, Oxford.